

\input harvmac
\input epsf

\def\figin{\epsfcheck\figin}\def\figins{\epsfcheck\figins}
\def\epsfcheck{\ifx\epsfbox\UnDeFiNeD
\message{(NO epsf.tex, FIGURES WILL BE IGNORED)}
\gdef\figin##1{\vskip2in}\gdef\figins##1{\hskip.5in}
\else\message{(FIGURES WILL BE INCLUDED)}%
\gdef\figin##1{##1}\gdef\figins##1{##1}\fi}
\def\DefWarn#1{}
\def\figinsert{\goodbreak\midinsert}
\def\ifig#1#2#3{\DefWarn#1\xdef#1{fig.~\the\figno}
\writedef{#1\leftbracket fig.\noexpand~\the\figno}%
\figinsert\figin{\centerline{#3}}\medskip\centerline{\vbox{\baselineskip12pt
\advance\hsize by -1truein\noindent\footnotefont{\bf Fig.~\the\figno:} #2}}
\bigskip\endinsert\global\advance\figno by1}

\lref\REFkazmig{V.A. Kazakov and A.A. Migdal, Nucl.Phys. {\bf B397}
(1993) 214}
\lref\REFmigd{A.A. Migdal, Mod.Phys. Lett. {\bf A8} (1993) 359}
\lref\REFmigtwo{A.A. Migdal, {\sl 1/$N$ Expansion and Particle Spectrum
in Induced QCD}, preprint PUPT-1332, July 1992}
\lref\REFg{D. Gross, Phys. Lett. {\bf B293} (1992) 181}
\lref\REFaoki{S. Aoki and A. Gocksch, {\sl The Spectrum of the
Kazakov--Migdal Model}, preprint UTHEP-250, December 1992}
\lref\REFizu{C.Itzykson and J.-B. Zuber, J. Math. Phys. {\bf 21} (1980) 411}
\lref\REFboul{D.V. Boulatov, {\sl Critical Scaling in the matrix
model on a Bethe tree}, preprint NBI-HE-93-55, September 1993}
\lref\REFm{A. Matytsin, Nucl.Phys. {\bf B411} (1994) 805}
\lref\REFluch{Luochi Chen, 1992, unpublished.}

\Title{\vbox{\baselineskip12pt\hbox{PUPT-1460}
\hbox{hep-th/9404096}}}
{\vbox{\centerline{Edge Singularity in ``Induced QCD"}\vskip0.15in
\centerline{}}}

\centerline{
\vbox{\hsize3in\centerline{Andrei Matytsin}
\smallskip\centerline{matytsin@puhep1.princeton.edu}
}}
\bigskip
\centerline{and}
\bigskip
\centerline{
\vbox{\hsize3in\centerline{Alexander A. Migdal}
\smallskip\centerline{migdal@math.princeton.edu}
}}
{\it
\bigskip\centerline{Department of Physics}
\centerline{Joseph Henry Laboratories}
\centerline{Princeton University}
\centerline{Princeton, NJ \ 08544}}
\vskip .5in

\noindent
The behaviour of the master field in ``induced QCD" near the edge
of its support is studied. An extended scaling domain, where the shape of
the master field is a universal function, is found.
This function is determined explicitly for the case of dimensions,
close to one, and the $D-1$-expansion is constructed.
The problem of the meson spectrum corresponding to this solution
is analyzed. As a byproduct of these calculations, a new, explicit
equation for the meson spectrum in ``induced QCD" with a general
potential is derived.

\Date{April 1994}

\newsec{Introduction.}

The ``induced QCD" is a lattice gauge theory of a scalar matrix field
$\Phi(x)$ defined by the functional integral \REFkazmig
\eqn\inducedQCD{\eqalign{
{\cal Z}= \int{\cal D}\Phi(x){\cal D}U_{\mu}(x) \thinspace
{\rm exp}\biggl\{-N\sum_{x}\Bigl[&{\rm tr} U\bigl(\Phi(x)\bigr) \cr
-&{\rm tr}\sum_{\mu}\Phi(x)U_{\mu}(x)\Phi(x+\mu a)
U_{\mu}^{\dagger}(x)\Bigr]\biggr\}.\cr}
}
In this formula $x$ marks the sites of the $D$-dimensional cubic lattice
with lattice spacing $a$, $\Phi(x)$ are $N\times N$ Hermitian matrices
and $U_{\mu}(x)$ is the gauge field ($\mu$~referring to the directions
of lattice links, $\mu= 1, \ldots, D$).

The field $\Phi(x)$ interacts with itself through the potential $U(\Phi)$
and has the obvious kinetic term
${\rm tr}\big[\Phi(x)U_{\mu}(x)\Phi(x+\mu a)U_{\mu}^{\dagger}(x)\big]$.
This term is the only place where the gauge field enters this theory.
Indeed, no special action for the gauge field is included in the model.

Since $U_{\mu}(x)$ is incorporated in the action in such a simple way,
the ``induced QCD" is, in principle, exactly solvable in the large
$N$ limit.
In this respect \inducedQCD\ is one of the very few
matrix theories which
admit an exact solution not only in $D\le1$, but
also in $D>1$ dimensions. Needless to say, it would be extremely interesting
to investigate such a solution, especially in the $D>1$ case.
In particular, the scaling
properties of this model near its phase transition point must exibit
universal
features, characteristic of other $D>1$ models.

In the large $N$ limit the eigenvalue density $\rho(\phi)$
of the field $\Phi(x)$
obeys classical equations of motion. That is to say, $\rho(\phi)$
plays the role of the master field in ``induced QCD". Its classical
dynamics can be formulated in two ways.
One way is to use the Schwinger--Dyson equations for the functional integral
\inducedQCD, obtaining a nonlinear singular integral equation
for $\rho(\phi)$ \REFmigd. Quite remarkably, for the case of quadratic
$U(\Phi)$ this equation can be solved exactly \REFg.
On the other hand, the same density $\rho(\phi)$
is related to solutions of a certain
quasilinear partial differential equation (the Hopf equation)\REFm.
These two formulations may appear very different, but in fact they are
equivalent. Under some general assumptions about the
potential $U(\Phi)$ both of them reduce to the same
functional equation \REFm, \REFluch, \REFboul.

In the vicinity of critical points, where $\rho(\phi)$ vanishes, the
small fluctuations of the eigenvalue density become significant,
leading to critical behaviour.
The location of the critical point depends on parameters of the potential,
but the critical indices are, presumably, universal.
Remarkably, in the critical domain the master field equation
simplifies, which allows one
to determine the critical indices without
constructing the complete solution of the whole master field
equation.

The critical point can be positioned either inside of the
support of $\rho(\phi)$ or at the edge of this support.
The critical indices, corresponding to these two cases,
are different.
If $\rho(\phi)$ vanishes inside
of its support (for an even potential this would occur at $\phi=0$)
the eigenvalue density behaves at the critical point as \REFmigd\
$$\rho(\phi)\propto |\phi|^{1+\gamma}, \qquad
\cos \pi\gamma = {D\over 3D-2}.$$
This case is rather sophisticated.
In particular, an important problem of
the mass spectrum \REFmigtwo\
corresponding to this critical point is still unsolved.

The other, simpler situation was considered
recently by Boulatov \REFboul.
He assumed that the master field vanishes at the endpoint
of its
support according to
$$\rho(\phi)\propto (\phi- a)^{1+\gamma} \theta(\phi- a).$$
He found two possible solutions for  $\gamma$:
$$\cos\pi\gamma=D \quad {\rm or} \quad \cos\pi\gamma={D\over 2D-1}.$$

However, it is still unclear how the Boulatov's master field behaves
away from the endpoint. The scalar field potential giving rise
to this solution is not known either. Finally, although,
by construction, this master field corresponds to an extremum
of the action of the ``induced QCD", it may maximize, rather than
minimize, the action.

In this paper we will take a closer look at Boulatov's solution.
In section 2
we determine the shape of the master field away from the
endpoint. We find that there is a whole domain where
this field is described by certain real universal functions,
which can be represented in terms of a power series.
In section 3 we consider the case of dimensions, close to 1,
and construct the $D-1$ -expansion of the master field.
This allows us to sum the series and evaluate these universal
functions explicitly.

Finally, we address the problem of
spectrum in ``induced QCD". In section 4 we  use the connection between the
``induced QCD" and the Hopf boundary problem to obtain a new,
simpler form of the wave equation, describing the meson spectrum of
the theory. This derivation applies to ``induced QCD" with any potential
and is not in any way restricted to Boulatov's solution,
the spectrum of which we shall determine in section 5. Our calculations
show that in this case the meson spectrum
contains tachyons. This implies that such solution does not correspond
to a local minimum of the action.  Nevertheless,
the master field we are investigating exibits some general
features which are likely to be present in other, healthier
solutions.

\newsec{The Series Expansion of the Master Field.}

The exact solution of the large-$N$ ``induced QCD" is based on the
saddle point method \REFkazmig, \REFmigd.
First, one integrates out $U_{\mu}(x)$ in
\inducedQCD\ using the Itzykson-Zuber integral \REFizu
\eqn\iz{
I(\Phi, \Psi)=\int{\cal D}U \thinspace
{\rm exp}[N{\rm tr}\Phi U \Psi U^{\dagger}]=
{{\rm det}\big[{\rm exp}(N\phi_i \psi_j)\big] \over \Delta(\phi) \Delta(\psi)}
,}
where
$$\Delta(x)=\prod_{i<j}(\phi_i-\phi_j),$$
$\phi_i$ and $\psi_j$ being the eigenvalues of the $N\times N$ matrices
$\Phi$ and $\Psi$. One is left then with an effective theory of the
``eigenvalue fields" $\phi_i (x)$ which has the action\foot{The
second term in $S_{\rm eff}$ is due to the fact that the integration
measure ${\cal D}\Phi(x)$ after the change of variables to $\phi_i(x)$
reduces to $\Delta^2 (\phi(x)) \prod_{i}{\cal D}\phi_i (x)$.}
\eqn\eff{\eqalign{
S_{\rm eff}[\phi(x)]=&N\sum_{x, i}U\bigl(\phi_i (x)\bigr)-\sum_{x, i\ne j}
{\rm ln}\bigl|\phi_i (x)-\phi_j (x)\bigr|\cr
&+\sum_{x, \mu}\ln\Bigl[I\bigl(\phi(x), \phi(x+\mu a)\bigr)\Bigr].\cr
}}
In the large $N$ limit the remaining functional integral over $\phi_i (x)$
is dominated by the saddle point of $S_{\rm eff}[\phi(x)]$, that is, by
$x$-independent
$\phi_i (x)=\phi_i$ satisfying
\eqn\sad{
{\partial\over \partial\phi_i}S_{\rm eff}[\phi]=0.
}

Since the effective action $S_{\rm eff}$ is invariant with respect to any
permutation of $\phi_i, i=1, \ldots, N$, the saddle point can be fully
described by specifying the density of eigenvalues $\phi_i$ on the real line,
$\rho(\phi)$. This density plays the role of the ``master field" in
this theory. To solve the ``induced QCD" means to find $\rho(\phi)$
for the given interaction potential $U(\Phi)$. This, in turn, allows one
to evaluate other physical observables of the theory.

As a consequence of the saddle point condition \sad, $\rho(\phi)$ obeys
a certain nonlinear singular integral equation \REFmigd.
However, for a generic
potential $U(\Phi)$ it is possible to derive a {\it functional} constraint
fixing $\rho(\phi)$ \REFm. One introduces the two
functions\foot{This formula defines $G_{\pm}(x)$ for {\it real} $x$. For
complex $x$, $G_{\pm}(x)$ are defined by
means of analytic continuation from the real
axis.}
\eqn\gpm{
G_{\pm}(x)={1\over 2D}U^{\prime}(x) +{D-1\over D}\thinspace
{\cal P}\!\int{\rho(y)dy
\over x-y} \pm i \pi \rho(x).
}
Then the saddle point of the ``induced QCD" is determined by the
equations\foot{It is not known, however, whether or not some nonanalytic
solutions of induced QCD can be described by this constraint.}
\eqn\gpgm{
\eqalign{ &G_+\bigl(G_-(x)\bigr)=x, \cr
	  &G_-\bigl(G_+(x)\bigr)=x. \cr
}}
That is to say, $G_+(x)$, as a function of $x$, is inverse with respect to
$G_-(x)$.

Generally speaking, the master field $\rho(x)$ is not equal to zero only
inside a finite interval (referred to as the support of $\rho(x)$),
and vanishes at its endpoints. However, it is the
behaviour of $\rho(x)$ in the vicinity of endpoints that determines the
universal properties of the theory. For example, if $\rho(x)\simeq
(a-x)^{\gamma+1}$ near the endpoint $a$, then
the exponent $\gamma$ is universal.

The general solution of equations \gpgm\ is unknown. Therefore, there is
no direct way to classify all possible values of $\gamma$ that can be
achieved in ``induced QCD". Nevertheless, in the vicinity of a point where
$\rho(x)$ vanishes, the equations should simplify. Boulatov
\REFboul\ has observed
that \gpgm\ admits a very simple solution, $G_+(x)=G_-(x)=-x$. Clearly,
such solution by itself does not describe the ``induced QCD" with any
potential, since it contains no imaginary part at all.
To correct this drawback,
Boulatov perturbed $G_\pm$ with a powerlike term,
\eqn\expansion{
G_+(x)= -x + \alpha (-x)^{1+\gamma}+\ldots.
}
If $\gamma$ is not an integer, then for $x>0$
$$\pi\rho(x)={\rm Im}\thinspace G_+(x)=
\alpha x^{1+\gamma} {\rm sin}(\pi \gamma) +\ldots \ne 0.$$
Boulatov found that this ansatz is consistent with \gpm\ only if
$\gamma$ assumes a certain value,
\eqn\gammaeq{
\cos\pi\gamma=D \quad {\rm or} \quad \cos\pi\gamma={D\over 2D-1}.
}
In other words, the behaviour of the master field near the endpoint of
its support is completely determined by the dimension $D$. Notice that,
except for some special cases, $\gamma$ is not a rational number.

Since the constraint \gpgm\ is nonlinear, any perturbation like
$(-x)^{1+\gamma}$ will inevitably generate perturbations of
higher orders. One ends up with a whole series for $G_+(x)$,
\eqn\gser{
G_+(x)=-x+\sum_{k=1}^{\infty}g_k(-x)^{\gamma k+1}.
}
If $\gamma$ were rational, $\gamma k +1$ could become an integer for
a sufficiently large $k$. In this case the $k$-th order term
should be interpreted as induced by the (polynomial) potential
$U(x)$ in \gpm. However, if $\gamma$ is not rational,
such phenomenon can never occur, and all of the coefficients $g_k$
can be determined dynamically, from the master field equations \gpgm\ alone.
This suggests that the whole infinite series \gser, not only its
first term $(-x)^{1+\gamma}$, has some universal properties.
Let us show that this is indeed the case.

To begin with, let us note that the function
\eqn\fofx{
f(x)={1\over 2(D-1)}U^{\prime}(x)+\int {\rho(y)dy\over x-y}
}
is an analytic function of $x$, which has a cut along the support of
$\rho(x)$. Since on the cut
$$f(x\pm i0)={1\over 2(D-1)}U^{\prime}(x)+
{\cal P}\!\int {\rho(y)dy\over x-y} \mp i\pi\rho(x),
$$
it follows from \gpm\ that
\eqn\fg{
\eqalign{
&G_+(x)=-x+{2D-1\over 2D}\overline{f}(x)-{1\over 2D}f(x),\cr
&G_-(x)=-x+{2D-1\over 2D}f(x)-{1\over 2D}\overline{f}(x).\cr
}}

Boulatov's ansatz \expansion\ essentially means that $f(x)$ has a branch
cut along the real axis from $0$ to $+\infty$.
Consequently, it can be expanded
as
\eqn\fexp{
f(x)=x\thinspace\sum_{k=1}^{\infty}f_k (-x)^{\gamma k}.
}
This expansion makes sence by itself for real negative $x$. To define
$f(x)$ for real positive $x$, we need to perform the analytic continuation,
with the result
$$f(x)=x\thinspace\sum_{k=1}^{\infty}f_k z^k x^{\gamma k}, $$
where
$$z={\rm e}^{i\pi\gamma}.$$
Then
$$\overline{f}(x)=x\thinspace\sum_{k=1}^{\infty}f_k z^{-k} x^{\gamma k}.$$
Substituting this into \fg\ and introducing the notation
\eqn\abdef{
\eqalign{
&A_k={z^k\over 2D} - {2D-1\over 2D}z^{-k}, \cr
&B_k={1\over 2D}- {2D-1\over 2D}z^{2k}, \cr
}}
we easily derive the following expansions:\par
---for real positive $x$ on the upper edge of the cut, ${\rm Im}\thinspace
x = +i0$:
\eqna\gpmpos
$$\eqalignno{&G_+(x)=-x\biggl(1+\sum_{k=1}^{\infty}
f_k A_k x^{\gamma k}\biggr)&\gpmpos a\cr
&G_-(x)=-x\biggl(1+\sum_{k=1}^{\infty}
f_k B_k z^{-k} x^{\gamma k}\biggr)&\gpmpos b\cr
}
$$
\par
---for real negative $x$ (there is no cut at $x<0$):
\eqna\gpmneg
$$\eqalignno{
&G_+(x)=-x\biggl(1+\sum_{k=1}^{\infty}
f_k A_k z^k (-x)^{\gamma k}\biggr)&\gpmneg a\cr
&G_-(x)=-x\biggl(1+\sum_{k=1}^{\infty}
f_k B_k (-x)^{\gamma k}\biggr)&\gpmneg b\cr
}
$$

Our goal is to determine $f_k$. To this end,
we must impose the constraint \gpgm. This can be done
in two different inequivalent ways. The first way is
to impose $G_+\bigl(G_-(x)\bigr)=x$ on the {\it upper\/}
edge of the cut, where
${\rm Im}\thinspace  x=+i0$.
\ifig\mappingsone{(left) The functional composition
$G_+\bigl(G_-(x)\bigr)$ takes a real point $x>0$ to itself. More precisely,
$G_+\bigl(G_-(x)\bigr)$ maps the upper edge of the branch cut onto the
lower edge.$\;$ (right) Both $G_+\bigl(G_-(x)\bigr)=x$ and
$G_-\bigl(G_+(x)\bigr)=x$ are satisfied on the negative part of the
real axis.}
{\epsfxsize 2.0in \epsfbox{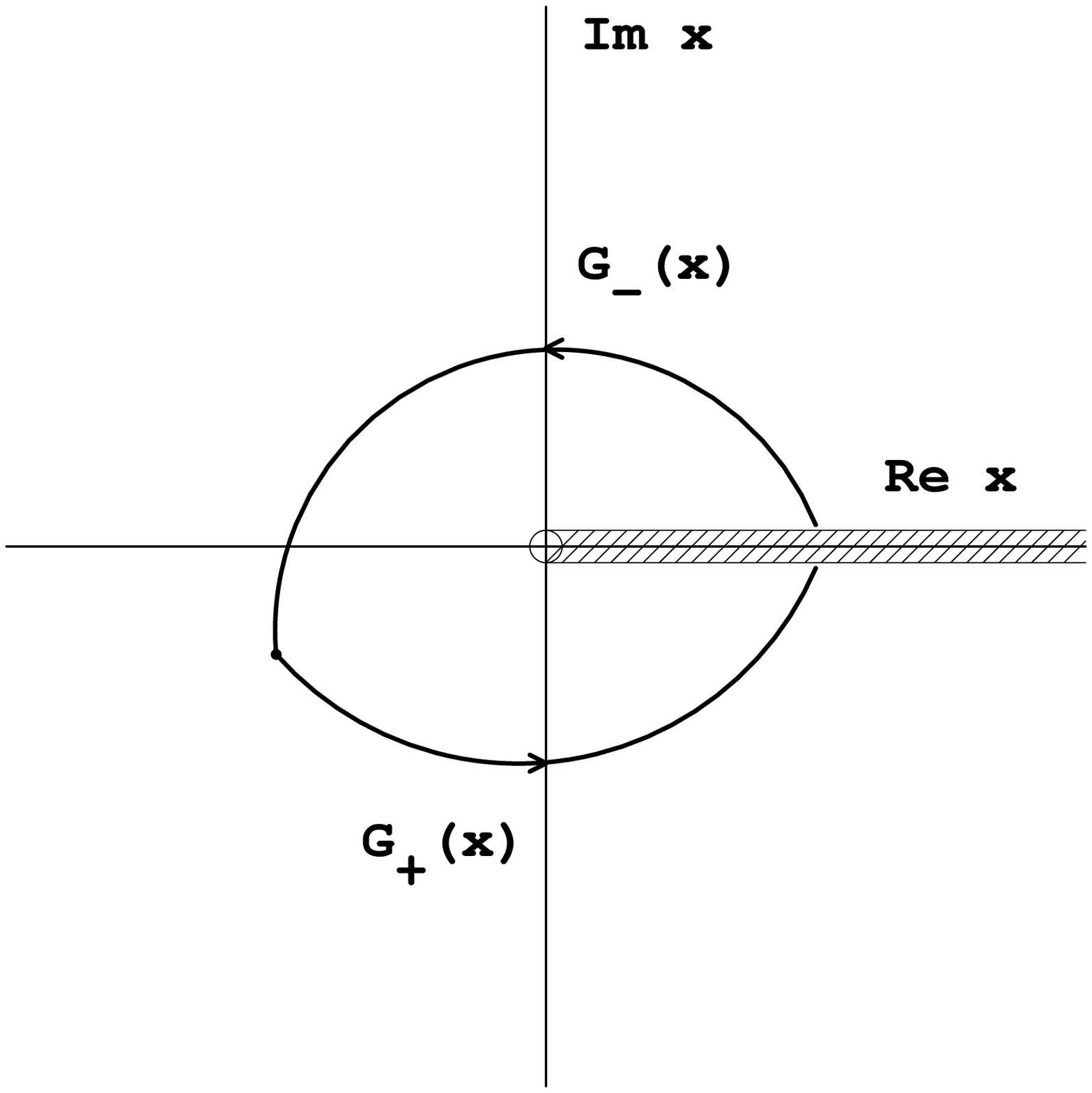}\hskip 0.5in
 \epsfxsize 2.0in \epsfbox{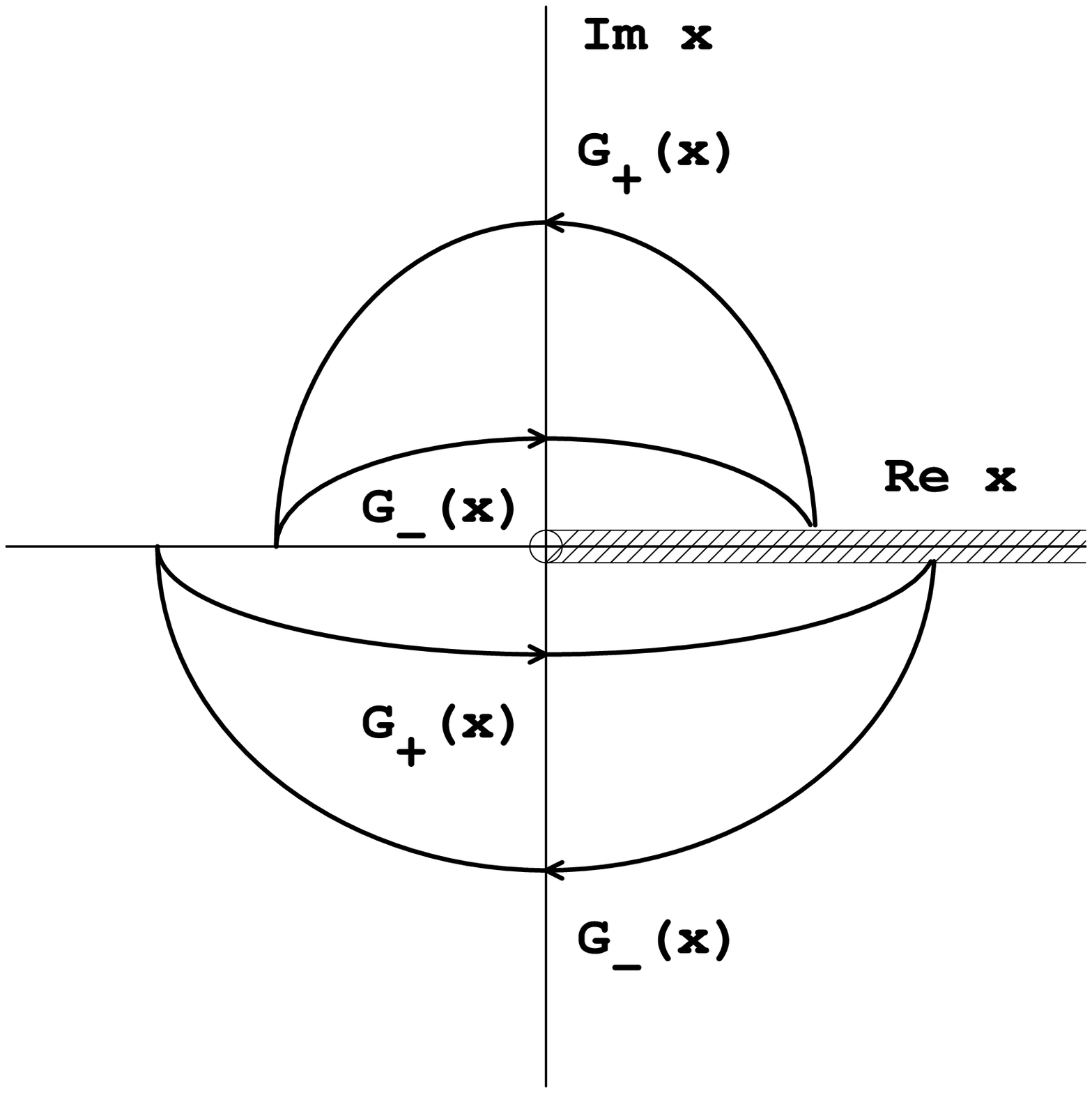}}
In view of \gpmpos{a}\ and \gpmneg{b}\ this requirement
can be written down as
\eqn\horrible{
\Bigl(1+\sum_{k=1}^{\infty}f_k A_k x^{\gamma k}\Bigr)\biggl(1+
\sum_{p=1}^{\infty}f_p B_p x^{\gamma p} \Bigl(
1+\sum_{q=1}^{\infty}f_q A_q x^{\gamma q}\Bigr)^{\gamma p}\biggr)=1.
}
Then one can check that the other constraint, $G_-\bigl(G_+(x)\bigr)=x$,
is fulfilled on the {\it lower\/} edge of the cut, where
${\rm Im}\thinspace x=-i0$. In
addition, {\it both\/} of the constraints \gpgm\ are satisfied on the real
axis {\it outside\/} of the cut, at $x<0$\thinspace\ (see \mappingsone).

The second way to impose \gpgm\ is to demand $G_+\bigl(G_-(x)\bigr)=x$ on
the lower edge of the cut. This will lead to an equation, different from
\horrible. As the treatment of these two cases is completely analogous,
we will now concentrate on the first one. After that, we will briefly
describe the second case.

Equation \horrible\ allows us to find the coefficients $f_k$
recursively, one after another. To do this, we expand the right hand side
of \horrible\ in a power series of a formal parameter $t\equiv x^{\gamma}$:
$$\eqalign{
1+ (A_1+B_1)f_1& t+ \bigl( (A_2+B_2)f_2 + (1+\gamma)A_1 B_1 f_1^2\bigr)t^2
\cr
+\Bigl[(A_3+B_3)f_3&+ \bigl((1+\gamma)A_2 B_1+ (1+2\gamma)A_1 B_2\bigr)
f_1 f_2 \cr
&+{\gamma (1+\gamma)\over 2}A_1^2 B_1 f_1^3\Bigr]t^3+\ldots =1.\cr
}
$$
Since $f_1\ne 0$, we obtain the consistency condition
\eqn\consist{
A_1+B_1=0
}
as well as
\eqna\fhigh
$$\eqalignno{
f_2=-(1+\gamma)&{A_1 B_1\over A_2+B_2}f_1^2, &\fhigh a\cr
f_3=-(1+\gamma)&{A_1 B_1\over(A_2+B_2)(A_3+B_3)}\Bigl((1+\gamma)(A_1B_2+
B_1A_2)\cr
&\qquad\qquad\qquad\qquad\qquad
+{\gamma\over 2}A_1(B_2-A_2)\Bigr)f_1^3, &\fhigh b\cr
&\ldots\cr
}$$
Using \abdef\ and \consist, we see that the consistency condition
translates into
\eqn\z{
z+{1\over z}={2D\over 2D-1}
}
which, in view of $z={\rm e}^{i\pi\gamma}$, implies the Boulatov's result
\gammaeq. Since $\gamma$ has to be real, this solution makes sense
only for $D>1$ or $D<1/3$.

We can simplify \fhigh{}, using \abdef\ to express $A_k$ and $B_k$
in terms of $z$, and eliminating $D$ by virtue of \z :
\eqn\fhighz{
\eqalign{
&f_2=-{1+\gamma\over z^{-1}+1+z}f_1^2
=-(1+\gamma){2D-1\over 4D-1}f_1^2, \cr
&f_3={(1+\gamma)(2+3\gamma)\over 2\bigl(z^{-2}+z^{-1}+2 +z+z^2\bigr)}
f_1^3
=(1+\gamma)(2+3\gamma){(2D-1)^2\over  4D(4D-1)}f_1^3, \cr
}}
and so on.
Note the remarkable feature of these expressions: they are all real.
Indeed, since $z$ is a complex number of absolute value $1$, the
denominators of $f_k$ are invariant under complex conjugation.
It is not at all obvious from \horrible\ that this should be the case.
Indeed, the reality of $f_k$ is closely connected to the fact that
outside of the cut the first of the two equations \gpgm,
$G_+\bigl(G_-(x)\bigr)=x$, implies the second, $G_-\bigl(G_+(x)\bigr)=x$,
and vice versa. Let us now prove that all of the $f_k$ are given by
real numbers.

We will do this in two steps. The statements below are
easy to infer by inspection of \fhigh{}.

{\it When viewed as functions of $\{A_1, A_2, \ldots\}$,
$\{B_1, B_2, \ldots\}$, the functions $f_k\bigl(\{A_i\}, \{B_j\}\bigr)$
are symmetric with respect to the interchange of $A$ and $B$:
$$f_k\bigl(\{A_i\}, \{B_j\}\bigr)=f_k\bigl(\{B_i\}, \{A_j\}\bigr),$$
provided that the consistency condition \consist\ is satisfied.}

At first sight, this might appear false. Indeed, the expression for $f_3$
in \fhigh{b}\ is not at all symmetric under the interchange of $A$ and $B$.
However, its asymmetric part equals
$$f_3\bigl(\{A_i\}, \{B_j\}\bigr)-f_3\bigl(\{B_i\}, \{A_j\}\bigr)=
{\gamma(1+\gamma)\over 2}{A_1 B_1\over A_3+B_3}{A_2-B_2\over A_2+B_2}
(A_1+B_1)$$
and vanishes due to the consistency condition \consist.

To prove the symmetry of $f_k$, we introduce an auxiliary variable
\eqn\expu{
u=x\Bigl(1+\sum_{k=1}^{\infty}f_k A_k x^{\gamma k}\Bigr).
}
Then it is a consequence of \horrible\ that
\eqn\expx{
x=u\Bigl(1+\sum_{p=1}^{\infty}f_p B_p u^{\gamma p}\Bigr).
}
Substituting \expx\ back into \expu, we get
\eqn\horr{
u=u\Bigl(1+\sum_{p=1}^{\infty}f_p B_p u^{\gamma p}\Bigr)\biggl(1+
\sum_{q=1}^{\infty}f_q A_q u^{\gamma q} \Bigl(
1+\sum_{l=1}^{\infty}f_l B_l u^{\gamma l}\Bigr)^{\gamma q}\biggr)
}
which is the same equation as \horrible, but with all $A_i$ replaced by
$B_i$ and vice versa. Since the $f_k$ can be recursively
determined from either of the
equations \horrible\ or \horr\ and since $f_k$ are unique, they have to
be symmetric in $A$ and $B$.

{\it If we assign to $A_j$ and $B_j$ a formal degree of $j$,
then $f_k$ has degree zero for any $k$.}

That is to say, if we scale
\eqn\lem{
\eqalign{
&A_i^{(\lambda)}=\lambda^i A_i, \cr
&B_i^{(\lambda)}=\lambda^i B_i, \cr
}}
then
$$f_k\bigl(\{A_i^{(\lambda)}\}, \{B_j^{(\lambda)}\}\bigr)=
f_k\bigl(\{A_i\}, \{B_j\}\bigr).$$

This statement follows immediately from the observation
that such rescaling amounts to a redefinition of expansion parameter $x$
in \horrible, $x^{\gamma}\rightarrow \lambda x^{\gamma}$, which does not
change the equations for $f$.

Now we are able to prove that $f_k$ is real for any $k$.
Indeed, we can eliminate $D$ in \abdef\ in terms of $z$,
using \z. This gives
\eqn\abz{\eqalign{
&A_k(z)=z^k- {z\over z^2+1}\Bigl(z^k+{1\over z^k}\Bigr),\cr
&B_k(z)=1-{z\over z^2+1}\bigl(1+z^{2k}\bigr).\cr
}}
Under complex conjugation,
$$\eqalign{
&A_k^*(z)=z^{-k}B_k(z),\cr
&B_k^*(z)=z^{-k}A_k(z).\cr
}$$
Therefore,
$$f_k^*\bigl(\{A_i\}, \{B_j\}\bigr)=f_k\bigl(\{A_i^*\}, \{B_j^*\}\bigr)
=f_k\bigl(\{z^{-i}B_i\}, \{z^{-j}A_j\}\bigr).$$
Setting $\lambda=z^{-1}$ in \lem,
$$f_k\bigl(\{z^{-i}B_i\}, \{z^{-j}A_j\}\bigr)=
f_k\bigl(\{B_i\}, \{A_j\}\bigr)$$
and, by the symmetry proprety,
$$f_k\bigl(\{B_i\}, \{A_j\}\bigr)=f_k\bigl(\{A_i\}, \{B_j\}\bigr).$$
Hence
$$f_k^*\bigl(\{A_i\}, \{B_j\}\bigr)=f_k\bigl(\{A_i\}, \{B_j\}\bigr),$$
so that all of the $f_k$ are indeed real.

One can see from \fhighz\ that $f_k/f_1^k$ are universal numbers, which
depend only on the dimension $D$. The coefficient $f_1$
(which can be arbitrary) enters all formulas in the combination
$f_1 x^{\gamma}$, thus setting the scale for $x$. We conclude
that there is a whole scaling domain, where $f_1 x^{\gamma}\sim 1$.
The (universal) behaviour of the master field in this domain is
more complicated than the powerlike shape of the singularity
at $x\rightarrow 0$. In the next section we will
investigate this behaviour in more detail.

Finally, let us note that the condition on $\gamma$, \z,
admits not only a positive, but also a negative solution,
$\gamma=-(1/\pi)\thinspace \arccos D/(2D-1)$.  In this case
\expansion\ is not a good approximation at all. However, if one takes
into account the whole series for $f(x^{\gamma})$, it is possible
to derive the behaviour of
$\rho(x)$ as $x\rightarrow 0$. In the next section
we will see that in this case the master field develops
a logarithmic singularity.

\newsec{The $D-1$-expansion.}

One might wonder how the universal function $f$ behaves at finite values
of its argument. Obviously, the first few terms of a Taylor series do not
answer this question. One needs to evaluate all coefficients of the series
and perform the summation.

This can be done explicitly when $\gamma\rightarrow 0$. Indeed, it is easy
to see from \fhighz\ that the $f_k(\gamma=0)$ are not singular.
We will see below that in fact they describe the first order
in the $D-1$-expansion around $D=1$.
To calculate the numbers $f_k\equiv f_k(\gamma=0)$, we return to the general
formula \horrible, with $t\equiv x^{\gamma}$:
\eqn\horrt{
\Bigl(1+\sum_{k=1}^{\infty}f_k A_k t^k\Bigr)\biggl(1+
\sum_{p=1}^{\infty}f_p B_p t^p \Bigl(
1+\sum_{q=1}^{\infty}f_q A_q t^q\Bigr)^{\gamma p}\biggr)=1.
}
Let us take the limit of this equation as $\gamma\rightarrow 0$,
while keeping $t$ fixed. Using \abz\ we derive
\eqn\abas{
\eqalign{
&A_k(z)=i\pi\gamma k -{\pi^2\gamma^2\over 2} +{\cal O}(\gamma^3),\cr
&B_k(z)=-i\pi\gamma k +\pi^2\gamma^2 \Bigl(k^2-{1\over 2}\Bigr)
+{\cal O}(\gamma^3).\cr
}}
Expanding \horrt\ up to ${\cal O}(\gamma^3)$ and keeping in mind that
$A_k(z)\sim B_k(z)\sim {\cal O}(\gamma)$, we get
$$\sum_{k=1}^{\infty}f_k(A_k+B_k)t^k +\sum_{k, p=1}^{\infty}f_k f_p
A_k B_p t^{k+p} +{\cal O}(\gamma^3)=0.$$
By virtue of \abas\ this means that the
$f_k(\gamma=0)$ satisfy the identity
$$\sum_{k=1}^{\infty}(k^2-1)f_k t^k +\sum_{k, p=1}^{\infty}k p f_k f_p
t^{k+p}=0$$
which translates into an ordinary differential equation for
$$f(t)\equiv \sum_{k=1}^{\infty}f_k t^k,$$
$$t{d\over dt}t{d\over dt}f(t)- f(t) +\Bigl(t{d\over dt}f(t)\Bigr)^2=0.$$
Introducing a new variable $\xi={\rm log}\thinspace t$ and denoting
${\dot f}\equiv df/d\xi$, we obtain
\eqn\difur{
{\ddot f} +{\dot f}^2 -f =0.
}
The solution of this equation is given by
\eqn\fexplicit{
{\rm \log}\thinspace t=
\int\limits_{}^{f(t)}\sqrt{{2\over 2 u + {\rm e}^{-2u} -1}}du
.}
Taking into account that $f(t)\sim f_1 t$ as
$t\rightarrow 0$, we finally obtain
\eqn\fffff{
\log {f_1 t\over f(t)} = \int\limits_{0}^{f(t)}\biggl\{
\sqrt{{2 u^2\over 2 u + {\rm e}^{-2u} -1}}-1\biggr\}{du\over u}.
}
In the same approximation the master field equals
\eqn\mf{\eqalign{
\pi\rho(x)=&{\rm Im}\thinspace G_+(x+i0)=-x\sum_{k=1}^{\infty}f_k t^k
{\rm Im}\thinspace A^k\cr
=&-\pi\gamma x \sum_{k=1}^{\infty}kf_k t^k
=-\pi\gamma x {\dot f}(t=x^{\gamma})\cr
=&-\pi\gamma x \sqrt{f(x^{\gamma})-
\half\bigl(1-{\rm e}^{-2f(x^{\gamma})}\bigr)}.\cr
}}
Note that we have fixed $x^{\gamma}$ to be finite, although $\gamma
\rightarrow 0$.
This means that in the scaling domain, where our expansion is valid,
$|{\rm log} x|\sim {\cal O}(1/\gamma)$.

Since $\rho(x)$ has to be positive, \mf\ implies that $f \gamma<0$.
Therefore, there are two possibilities:
either $\gamma<0$ and $f>0$,
or $\gamma>0$ and $f<0$. In the second case the terms in the series
expansion of $f(t)$ get smaller and smaller, as $x\rightarrow 0$,
and Boulatov's result gives a good approximation. This picture
fails, however, in the first case, when $\gamma<0$. Then,
as $x\rightarrow 0$, we have to use the $t=x^{\gamma}\rightarrow\infty$
asymptotics for $f(t)$. From \fexplicit\ we derive, remembering
that $f>0$,
$${\rm log}\thinspace t \simeq \int\limits_{0}^{f(t)}{du\over \sqrt{u}}=
2\sqrt{f(t)}$$
so that
$$f(t)\simeq {1\over 4}{\rm log}^2 t, $$
and
$$\rho(x)\simeq-{\gamma^2\over 2}x\thinspace{\rm log}\thinspace x,
\qquad x\rightarrow 0.$$
We see that in the scaling regime with $\gamma<0$ the master field
develops a logarithmic singularity at the endpoint of its support.

Since as $D \rightarrow 1$,
$$\gamma=\pm {1\over \pi}\sqrt{2(D-1)},$$
\mf\ represents the leading order of the $\sqrt{D-1}$-expansion
around $D=1$. Note that the expansion parameter is $\sqrt{D-1}$,
rather than $D-1$. This is related to the fact that
at $D-1$ the master field vanishes identically.
The perturbation, therefore, is singular in $D-1$.

Higher orders of expansion can be obtained in a similar way.
To evaluate the first correction, we set
$f_k=f_k(\gamma=0)+\gamma f_k^{(1)}+\ldots$ and expand \horrible\ to
${\cal O}(\gamma^4)$.
After some algebra one derives that the function
$$f^{(1)}(t)\equiv \sum_{k=1}^{\infty}f_k^{(1)} t^k$$
satisfies an ordinary differential equation:
$${\ddot f^{(1)}}+ 2 f {\dot f^{(1)}}- f^{(1)} = -{\dot f}{\ddot f},$$
where ${\dot f^{(1)}}\equiv df^{(1)}/d\xi$,\ $\xi={\rm log}\thinspace t$, and
$f$ is the leading order result given by \fexplicit.

In fact, this $D-1$-expansion gives a quite accurate approximation.
We have calculated the function $f(t)$ numerically for $D=2$ by the following
extrapolation technique.  We used \fffff\ as a guide, and modified it as
follows
\eqn\fextrapolate{
\log {f_1 t\over f(t)} = \int\limits_{0}^{f(t)}\biggl\{
\sqrt{{2 u^2\over 2 u + {\rm e}^{-2u} -1}}g(u)-1\biggr\}{du\over u}.
}
The function $g(u) $ with $g(0) = 1$ was computed as a Taylor series from exact
equation at $ D \neq 1 $. Then, the Pad\'e approximation was used. We checked
the precision by comparing diagonal with nondiagonal approximants. The $15/15$
and $14/16$ approximant practically coinsided for $ -2 < u < 2 $. This is where
we numerically computed the corresponding integral, with $15/15$ approximant
for $ g(u) $.

The result, along with the plot of the function $f$ for
$\gamma =0$ is presented in the figure. We checked
numerically that the difference remains small even at
$D=3$.

\ifig\figs{Plots of the function $f(t)$ for $D\rightarrow 1$
($\gamma=0$) and for $D=2$.}
{\epsfxsize 3.5in \epsfbox{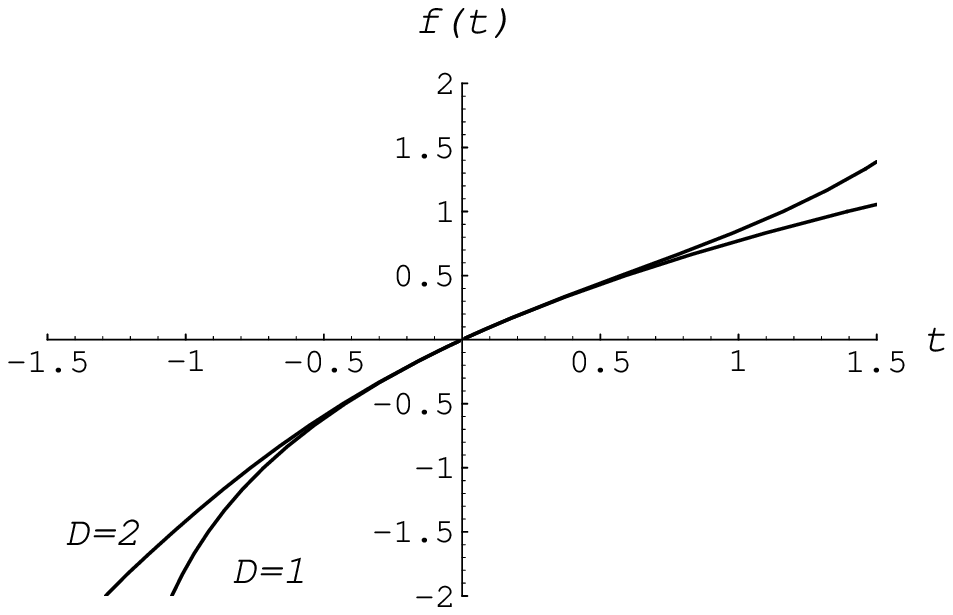}}
Analogous calculations can be carried out for the second solution,
which is obtained by imposing the constraint $G_+\bigl(G_-(x)\bigr)=x$
on the lower edge of the branch cut (see the discussion after
\horrible). This leads to another solution, valid at $|D|<1$,
\eqn\znew{
z+{1\over z}=2D,
}
It is possible to construct $1-D$-expansion
(in contrast with $D-1$-expansion we discussed
above), with $\gamma=\pm \sqrt{2(1-D)}/\pi$, \ $\pi \rho(x)=-\pi
\gamma {\dot{\tilde f}}(x^{\gamma})$, and
\eqn\fexplicitnew{
{\rm \log}\thinspace t=
\int\limits_{}^{{\tilde f}(t)}\sqrt{{2\over {\rm e}^{2u}-2u -1}}du
.}
In this case, if $\gamma<0$,\  ${\tilde f}(t)$ is defined only for
$f_1 t<1.27635$, meaning that the ansatz \fexp\ is internally
inconsistent. The only self-consistent possibility in this case is
$\gamma>0$, \ ${\tilde f}<0$.

In fact, the universal function defined by \fexplicitnew\ is just
a continuation of the function $f$ in \fffff\ to $t<0$. Indeed, using
\fffff\ and \fexplicitnew, and
keeping in mind that ${\tilde f}(t)\sim f(t)\sim f_1 t$ as
$ t\rightarrow 0$, it is easy to check that
$${\tilde f}(t)=-f(-t).$$
We see that the case $D>1$ is described by the function $f(t)$
with $t>0$, while its analytic continuation to $t<0$ will give
the solution for $D<1$.

If any of these solutions is to describe the ``induced QCD",
it has to minimize the action in \inducedQCD. However,
the master field equations  only
guarantee that the solutions above are the extremal points of the action.
To find the type of extremum corresponding to these solutions
one has to consider the quadratic variation of the action
functional and determine the normal modes and eigenfrequencies of
small oscillations around the master field.

\newsec{General Theory of the Spectrum in ``Induced QCD".}

In this section we will obtain an equation describing the normal modes of
small fluctuations around the master field in the
``induced QCD". An integral equation answering
this question has been derived before \REFmigtwo. However,
we will derive a new, simpler, equation in a different way. The result will
be applicable to the ``induced QCD" with any potential $U(\phi)$. We will
apply it to the particular case of the Boulatov's solution in the next
section.

As one can see from \eff, the effective
action of ``induced QCD" in the large $N$ limit
is
\eqn\effac{\eqalign{
S_{\rm eff}=N^2 \sum_{x}\biggl\{&\int \rho_x(\phi)U(\phi)d\phi-
\sum_{\mu}F\big[\rho_x(\phi), \rho_{x+\mu a}(\phi)\big]\cr
-&\int\int d\phi_1 d\phi_2 \rho_x(\phi_1)\rho_x(\phi_2){\rm ln}|\phi_1-
\phi_2|\biggr\}\cr}
}
where $\rho_x(\phi)$ is the eigenvalue density at the lattice site $x$,
and $F\big[\rho(\phi), \sigma(\psi)\big]$ is the large $N$ asymptotics
of the Itzykson--Zuber integral:
$$F\big[\rho(\phi), \sigma(\psi)\big]=
\lim\limits_{N\rightarrow\infty}\thinspace
{1\over N^2}\ln I[\Phi, \Psi]$$
$\rho(\phi)$ and $\sigma(\psi)$ being the eigenvalue distributions of
matrices $\Phi$ and $\Psi$.

We would like to see whether or not the small perturbations of the
master field are unstable.
To this end, let us derive the equation of motion for these
small perturbations (we shall refer to it as the wave equation).

Writing
$$\rho_x(\phi)=\rho_0(\phi)+\delta\rho_x(\phi)=
\rho_0(\phi)+\sum_{\vec p}
{\rm e}^{i{\vec p}{\vec x}}\delta\rho_{\vec p}(\phi)$$
where $\rho_0(\phi)$ is the translationally invariant master field,
determined from \sad, we find that the part of the action,
linear in $\delta\rho_{\vec p}(\phi)$, vanishes due to the master field
equation \sad,  while the part
quadratic in the fluctuations $\delta\rho_{\vec p}(\phi)$ is
\eqn\deltwos{
\delta_2 S_{\rm eff}=-\sum_{\vec p}\int d\phi_1 d\phi_2
\delta\rho_{\vec p}(\phi_1)K_{\vec p}(\phi_1, \phi_2)
\delta\rho_{-{\vec p}}(\phi_2)
}
where
\eqn\kernel{\eqalign{
K_{\vec p}(\phi_1, \phi_2) =&\ln |\phi_1-\phi_2|+
D{\delta^2 F[\rho, \sigma]\over \delta \rho(\phi_1) \delta \rho(\phi_2)}
\Biggr|_{\rho(x)=\sigma(x)=\rho_0(x)}\cr
&+\Omega^2({\vec p}){\delta^2 F[\rho, \sigma]\over \delta \sigma
(\phi_1) \delta \rho(\phi_2)}
\Biggr|_{\rho(x)=\sigma(x)=\rho_0(x)},\cr}
}
$$\Omega^2({\vec p})\equiv \sum_{\mu=1}^D \cos ap_{\mu}.$$
Note that since $\rho(x)=\sigma(x)$, the kernel $K_{\vec p}$ is symmetric
in $\phi_1$,\  $\phi_2$.

Since $\int\rho(\phi)d\phi$ always equals $1$, the perturbations
$\delta\rho_{\vec p}(\phi)$ must satisfy the normalization condition
$$\int\delta\rho_{\vec p}(\phi)d\phi=0.$$
Therefore, we can introduce the functions $\psi_{\vec p}(\phi)$ such that
$$\delta\rho_{\vec p}(\phi) ={d\over d\phi}\psi_{\vec p}(\phi)\thinspace,
\qquad \psi_{\vec p}(-\infty)=\psi_{\vec p}(+\infty)=0.$$
The equations of motion for $\psi_{\vec p}(\phi)$ can be obtained by
varying $\delta_2 S_{\rm eff}$ with respect to $\psi_{\vec p}(\phi)$:
\eqn\wave{
{\delta_2 S_{\rm eff}\over \delta \psi_{\vec p}(x)}=
2\int{\partial\over \partial x}K_{\vec p}(x, y)
\delta \rho_{\vec p}(y)dy=0.
}
This equation has solutions only for special discrete values of $\Omega_n^2$,
which form the spectrum of ``induced QCD".

To proceed, we will need to evaluate the second derivatives of the
Itzykson--Zuber integral, which are present in the kernel \kernel.
Although $I(\Phi, \Psi)$ is in general given by a simple explicit formula
\iz, the function $F[\rho, \sigma]$ has a very complicated structure.

Let us first outline the calculation of the first functional derivatives
of the Itzykson--Zuber integral.
It is convenient to introduce the two functions $v(\phi)$ and
$u(\phi)$, defined by
\eqn\deriv{
\eqalign{
&{\partial\over \partial\phi}{\delta F\big[\rho(\phi), \sigma(\psi)\big]
\over \delta \rho(\phi)}=\phi +v(\phi)-
{\cal P}\int{\rho(\phi^{\prime})d \phi^{\prime}\over \phi -\phi^{\prime}}, \cr
&{\partial\over \partial\psi}{\delta F\big[\rho(\phi), \sigma(\psi)\big]
\over \delta \sigma(\psi)}=\psi+u(\psi)-
{\cal P}\int{\sigma(\psi^{\prime})d\psi^{\prime}\over \psi-\psi^{\prime}}.\cr
}}
Then $v(\phi)$ and $u(\psi)$ can be determined as follows \REFm.
One introduces an auxiliary  complex function
$\varphi(x, t)$ and sets up the
boundary problem for the so-called Hopf equation:
\eqn\hopf{
\eqalign{
&{\partial \varphi(x, t)\over \partial t}+
\varphi(x, t){\partial \varphi(x, t)\over \partial x}=0, \cr
&{\rm Im}\thinspace \varphi(x, t=0)=\pi\rho(x), \cr
&{\rm Im}\thinspace \varphi(x, t=1)=\pi\sigma(x).\cr
}}
Then it is possible to prove that
$$\eqalign{
&v(x)=+{\rm Re}\thinspace \varphi(x, t=0), \cr
&u(x)=-{\rm Re}\thinspace \varphi(x, t=1).\cr
}$$
Let us emphasize that that we are considering the case when
the two eigenvalue distributions entering
the Itzykson--Zuber integral are not necessarily the same.

In terms of $v(x)$ and $u(x)$ the kernel of the wave equation becomes
\eqn\kerntwo{
{\partial\over \partial \phi_1}K_{\vec p}(\phi_1, \phi_2)=
(1-D){1\over \phi_1-\phi_2}+D{\delta v(\phi_1)\over \delta \rho(\phi_2)}
+\Omega^2({\vec p}) {\delta u(\phi_1)\over \delta \rho(\phi_2)}.}
To find the functional derivatives $\delta v(\phi_1)/\delta \rho(\phi_2)$
and $\delta u(\phi_1)/\delta \rho(\phi_2)$, which correspond to the second
variations of the Itzykson--Zuber integral,
we have to see how the solution of the Hopf equation \hopf\
changes when we vary the boundary conditions. This will
allow us to derive two integral equations, constraining
$\delta v(\phi_1)/\delta \rho(\phi_2)$ and
$\delta u(\phi_1)/\delta \rho(\phi_2)$. The easiest way
to do this is to use the fact that the Hopf equation
has infinitely many integrals of motion. We can choose
\eqn\integrals{
H_q(t)=-{1\over \pi t(q+1)}{\rm Im}\int dx
\bigl(x-t\varphi(x, t)\bigr)^{q+1}.
}
It is a consequence of \hopf\ that $dH_q/dt=0$. Hence
$H_q(t=1)=H_q(t=0)$, which, in view of the boundary conditions for
$\varphi$ entails
$$-{1\over \pi (q+1)}{\rm Im}\int dx \bigl(x-\varphi(x,1)\bigr)^{q+1}=
\int dx x^q \rho(x).$$
Considering a small variation of $\rho(x)$ and setting $\delta\sigma(x)=0$
(that is, varying one of the distributions in the Itzykson--Zuber integral
while keeping the other fixed),
we get
$$-{1\over \pi}{\rm Im}\int dx \delta u(x) \bigl(x+u(x)-i\pi\sigma(x)\bigr)^q
=\int dx x^q \delta \rho(x).$$
Summing these equations for $q=0, 1, \ldots$ with weight $z^{-q-1}$,
we obtain, provided that $z$ is outside of the support of $\rho(x)$:
$$-{1\over \pi}{\rm Im}\int dx {\delta u(x) \over
z-\bigl(x+u(x)\bigr)+i\pi\sigma(x)}=
\int dx {\delta\rho(x)\over z-x}.$$
Now we can put $\sigma(x)=\rho(x)=\rho_0(x), \ u(x)=v(x)$, and,
denoting $R(x)=x+u(x)=x+v(x)$, we obtain the first integral equation, which
determines $\delta u(x)/\delta \rho(y)$:
\eqn\varu{{\partial\over \partial y}
\int dx \biggl({\delta u(x)\over \delta \rho(y)}\biggr){\pi\rho(x)\over
\bigl(z-R(x)\bigr)^2 +\pi^2 \rho^2(x)}= {\partial\over \partial y}
{\pi\over z-y}.
}
To obtain an equation for $\delta v(x)/\delta \rho(y)$, we use
another set of
conservation laws
$$I_q(t)={1\over \pi (1-t)(q+1)}{\rm Im}\int dx
\bigl(x+(1-t)\varphi(x, t)\bigr)^{q+1}.
$$
Since now
$$\delta I_q(t=1)= \int dx x^q \delta\sigma(x)=0, $$
the same procedure yields
$$\delta I_q(t=0)= {1\over \pi}{\rm Im}\int dx \bigl(\delta v(x)
+i\pi\delta \rho(x)\bigr)\bigl(x+v(x)+i\pi\rho(x)\bigr)^q=0,$$
so that
$${\rm Im}\int dx {\delta v(x)
+i\pi\delta \rho(x)\over z-R(x)-i\pi\rho(x)}=0$$
and
\eqn\varv{
{\partial\over \partial y}
\int dx \biggl({\delta v(x)\over \delta \rho(y)}\biggr){\pi\rho(x)\over
\bigl(z-R(x)\bigr)^2 +\pi^2 \rho^2(x)}=-{\partial\over \partial y}
{z-R(y)\over \bigl(z-R(y)\bigr)^2 +\pi^2 \rho^2(y)}.
}
Notice that the integration weight in \varu\ and \varv\ is the same,
$${\pi\rho(x)\over
\bigl(z-R(x)\bigr)^2 +\pi^2 \rho^2(x)}.$$
So we can integrate the wave equation \wave\ over $x$ with this weight
to eliminate $\delta v(\phi_1)/\delta \rho(\phi_2)$ and
$\delta u(\phi_1)/\delta \rho(\phi_2)$.
Keeping in mind \kerntwo, we get the final form of the wave
equation\foot{We remind that, by construction, $z$ is positioned outside of
the support of $\rho(x)$.}
\eqn\wavefinal{\eqalign{
\int\Biggl[-{\Omega^2({\vec p})\over z-y}+&D\thinspace {z-R(y)\over
\bigl(z-R(y)\bigr)^2 +\pi^2 \rho^2(y)}\cr
+&(D-1)\thinspace{\cal P}\!\int {dx\over x-y}{\pi\rho(x)\over
\bigl(z-R(x)\bigr)^2 +\pi^2 \rho^2(x)}
\Biggr]\delta\rho_{\vec p}(y) dy=0.\cr
}}
The functions $R(x)$ and $\rho(x)$ are known from the solution
of the master field equation.
In fact, using \deriv, \hopf\ and \gpm, one can deduce that
$$G_{\pm}(x)=R(x)\pm i\pi\rho(x).$$

The equation we derived here is very explicit. It contains no auxiliary
functions and can be written down at once for any given solution of the
master field equation. For example, it can be used to derive the spectrum
of the ``induced QCD" with quadratic potential \REFg, recovering the
result of Aoki and Gocksch \REFaoki. We will use this equation to analyze
the Boulatov's spectrum.

\newsec{The Spectrum of the Solution with the Endpoint Singularity.}

The functions $R(x)$ and $\pi\rho(x)$ in the Boulatov's case equal
(see \expansion):
$$R(x)=-x-{\tilde \alpha}\thinspace
 {\rm cos}\thinspace \pi\gamma \thinspace x^{1+\gamma}
+\ldots,
\qquad \pi\rho(x)={\tilde \alpha}\thinspace
 {\rm sin}\thinspace \pi\gamma \thinspace
x^{1+\gamma}+\ldots.
$$
As $x\rightarrow 0$, we can approximate the weight in \wavefinal:
$${\pi\rho(x)\over
\bigl(z-R(x)\bigr)^2 +\pi^2 \rho^2(x)}
\rightarrow \delta\bigl(z-R(x)\bigr)\simeq \delta(z+x).$$
In this approximation the principal value
integral, present in the wave equation,
can be estimated as follows
$${\cal P}\!\int {dx\over x-y}{\pi\rho(x)\over
\bigl(z-R(x)\bigr)^2 +\pi^2 \rho^2(x)}= {\cal P}\!\int {dx\over
x-y} \delta(z+x) =-{1\over z+y}.$$
With the same accuracy,
$${z-R(y)\over
\bigl(z-R(y)\bigr)^2 +\pi^2 \rho^2(y)}\rightarrow
{\cal P}{1\over z-R(y)}\simeq {\cal P}{1\over z+y},$$
and we see that \wavefinal\ in the leading order becomes
\eqn\intlead{
\int\biggl[-{\Omega^2({\vec p})\over z-y} +{\cal P}{1\over z+y}\biggr]
\delta\rho_{\vec p}(y) dy=0.
}
This equation admits powerlike solutions, $\delta\rho_{\vec p}(y)=
y^{\alpha}$.
To find $\Omega^2({\vec p})$, we use the formulas\foot{We are keeping in mind
that $z<0$.}
\eqn\ints{\eqalign{
{\cal P}\!\int{y^{\alpha}dy\over z+y}&=-\pi {\rm cot}\thinspace \pi
\alpha \thinspace (-z)^{\alpha},\cr
\int{y^{\alpha}dy\over z-y}&={\pi\over {\rm sin}\thinspace \pi\alpha}
(-z)^{\alpha}.\cr
}}
This implies $$\Omega^2({\vec p})=- {\rm cos}\thinspace\pi\alpha.$$
At small lattice spacings
$$\Omega^2({\vec p}) =\sum_{\mu=1}^D {\rm cos}\thinspace a p_{\mu}
=D-{1\over 2} a^2 {\vec P}^2.$$
We see that, if $D>1$, whatever the index $\alpha$ is, the particles of
this theory have  ${\vec P}^2=-{\rm m}^2>0$. Hence, ${\rm m}^2<0$,
that is, these particles are tachyons.

To study the $D<1$ case we have to determine the index $\alpha$.
We shall argue that $\alpha=1+n\gamma$, with $n$ an integer.

Indeed, the wave equation \wave\ means that, as long as $\rho_0(\phi)$
solves the master field equation, so does $\rho_0(\phi)+
\delta\rho_x(\phi)$. Since $\rho_0(\phi)$ has an expansion in powers of
$\phi^{\gamma k}$, it is natural to look for small, coordinate-dependent,
perturbations of the coefficients of this expansion. This suggests
that the $n$-th normal mode of the of the wave equation arises
if we perturb the $n$-th coefficient in, say, \gser.
Such mode will be given by
$$\delta^{(n)} \rho_{\vec p}(x)=x^{1+\gamma n}\sum_{k=0}^{\infty}
d_k^{(n)}({\vec p})\thinspace x^{\gamma k}.$$

The functions $\delta^{(n)} \rho_{\vec p}(x)$ can be obtained by
linear combinations from the infinite set
$\{x^{1+\gamma}, x^{1+2\gamma}, \ldots\}$. A very important property
of this set is that it remains invariant under the action of
the wave operator $\partial {\hat K}/ \partial x$.

Let us represent the action of the integral operator in \wavefinal\
by an infinite-dimensional matrix $\hat M$:
$${
-\int{\pi\rho(x)dx\over
\bigl(z-R(x)\bigr)^2 +\pi^2 \rho^2(x)}
\int dy {\partial K(x, y)\over\partial x} \left\lgroup\! {\matrix{
y^{1+\gamma}\cr y^{1+2\gamma}\cr
\vdots\cr}}\!\right\rgroup = {\hat M}(\Omega^2)\left\lgroup\! {\matrix{
(-z)^{1+\gamma}\cr (-z)^{1+2\gamma}\cr
\vdots\cr}}\!\right\rgroup.}$$

The matrix ${\hat M}$ has an upper-triangular form
$${\hat M}=\pmatrix{\lambda_1(\Omega^2)& m^1_2 & m^1_3 & \ldots \cr
0&\lambda_2(\Omega^2)& m^2_3 &\ldots\cr
0& 0&\ddots & \ldots \cr
0& \vdots &\vdots &\ddots \cr}.
$$
The numbers $\lambda_i$ are the exact eigenvalues of ${\hat M}$.
Remarkably, they are determined  by the part of
$\partial {\hat K}/ \partial x$, which does not increase the
degree in $x$, that is, by the operator on the left hand side of
\intlead:
\eqn\lam{
\lambda_n(\Omega^2)=-{\pi\over {\rm sin}\thinspace \pi\alpha_n}
(\Omega^2+ {\rm cos}\thinspace \pi\alpha_n),
}
where $\alpha_n=1+n \gamma$. By adequately adjusting $\Omega^2$,
we can set any given $\lambda_n$ to zero, which would correspond to
the solution of the wave equation. Therefore,
the spectrum is given by
\eqn\spec{
\Omega^2_n=-{\rm cos}\thinspace \pi\alpha_n=
{\rm cos}\thinspace \pi n \gamma,}
so that $${\vec P}^2\sim D- {\rm cos}\thinspace \pi n \gamma.$$

If $D<1$, then $D={\rm cos}\thinspace \pi\gamma$, and
${\vec P}^2\sim{\rm cos}\thinspace \pi\gamma-
{\rm cos}\thinspace \pi n \gamma$. We see that, for $\gamma$
irrational, $-{\rm cos}\thinspace \pi n \gamma$ can
get arbitrarily close to 1,
thus again providing the evidence for tachyons in the spectrum.

The only possible loophole in this argument could be the locality of our
analysis. We did not find the global solution, but rather expanded  it near the
edge singularity.

In fact, we can decide whether our solution corresponds to a
minimum by checking that the eigenfrequencies of all normal
modes are real and not imaginary for any value of $\Omega^2$.
On the other hand, it is easy to see that the squares of
eigenfrequencies of the operator
$-{\partial {\hat K}/ \partial x}$
have the same sign as $\lambda(\Omega^2)$. If
$\alpha_n=1+n\gamma$,
$$\lambda_n(\Omega^2)={\pi\over {\rm sin}\thinspace \pi n \gamma}
(\Omega^2- {\rm cos}\thinspace \pi n \gamma).$$
Consequently, for small $n$ these are always positive
and do not cause any instability. However, at large enough $n \gamma > 1 $,
when $ \sin \pi n \gamma < 0 $ the sign in front of $P^2 $ changes,
which creates ghosts.

Although the higher
$n$ might indeed bring in ghosts and/or tachyons, one can imagine the
situation where these modes are excluded by the boundary
conditions. Strictly speaking, it is necessary to know the
master field throughout the whole region of support to
determine the spectrum fully.

\newsec{Conclusions.}

We have found that there is a special scaling domain where the
shape of the master field in ``induced QCD" is a universal function.
We have demonstrated how to calculate this function in terms of a
power series. Furthermore, we have constructed the
$D-1$-expansion to investigate the global structure of this
power series. Finally, we have evaluated the meson spectrum,
corresponding to this master field.

Independently of the above, we have used the description of
``induced QCD" in terms of the Hopf equation to derive a new, explicit,
version of the wave equation for the meson spectrum of the theory.
The equation we have found is valid for ``induced QCD" with any
potential and is not restricted to the particular application
we have considered in this paper.

\listrefs

\end